\documentclass[twocolumn,twocolappendix]{aastex631}
\usepackage{graphicx}
\usepackage{amsmath}
\usepackage{gensymb}
\usepackage{float}

\newcommand{\msun}{$\mathrm{M_{\odot}}$}
\newcommand{\rsun}{$\mathrm{R_{\odot}}$}
\newcommand{\lsun}{L$_{\odot}$}

\begin{document}
\title[24ggi progenitor]{The Red Supergiant Progenitor of Type II Supernova 2024ggi}

\correspondingauthor{Xiaofeng Wang}

\author[0000-0002-1089-1519]{Danfeng Xiang}\thanks{These authors contributed equally to this work.}
\affiliation{Department of Physics, Tsinghua University, Beijing 100084, China}
\author{Jun Mo}\thanks{These authors contributed equally to this work.}
\affiliation{Department of Physics, Tsinghua University, Beijing 100084, China}

\author[0000-0002-7334-2357]{Xiaofeng Wang}
\thanks{E-mail: wang\_xf@mail.tsinghua.edu.cn}
\affiliation{Department of Physics, Tsinghua University, Beijing 100084, China}
\author[0000-0002-1094-3817]{Lingzhi Wang}
\affiliation{Chinese Academy of Sciences, South America Center for Astronomy (CASSACA), National Astronomical Observatories, CAS, Beijing 100101, China}
\author[0000-0002-8296-2590]{Jujia Zhang}
\affiliation{Yunnan Observatories (YNAO), Chinese Academy of Sciences (CAS), Kunming, 650216, China}
\affiliation{International Centre of Supernovae, Yunnan Key Laboratory, Kunming 650216, China}
\affiliation{Key Laboratory for the Structure and Evolution of Celestial Objects, CAS, Kunming, 650216, China}

\author{Han Lin}
\affiliation{Yunnan Observatories (YNAO), Chinese Academy of Sciences (CAS), Kunming, 650216, China}
\affiliation{International Centre of Supernovae, Yunnan Key Laboratory, Kunming 650216, China}
\affiliation{Key Laboratory for the Structure and Evolution of Celestial Objects, CAS, Kunming, 650216, China}

\author{Liyang Chen}
\affiliation{Department of Physics, Tsinghua University, Beijing 100084, China}

\author[0000-0001-8390-9962]{Cuiying Song}
\affiliation{Department of Physics, Tsinghua University, Beijing 100084, China}

\author[0000-0002-8708-0597]{Liang-Duan Liu}
\affiliation{Institute of Astrophysics, Central China Normal University, Wuhan 430079, China}
\affiliation{Key Laboratory of Quark and Lepton Physics (Central China Normal University), Ministry of Education, Wuhan 430079, China}

\author{Zhenyu Wang}
\affiliation{Yunnan Observatories (YNAO), Chinese Academy of Sciences (CAS), Kunming, 650216, China}
\affiliation{International Centre of Supernovae, Yunnan Key Laboratory, Kunming 650216, China}
\affiliation{University of Chinese Academy of Sciences, Beijing 100049, China}

\author{Gaici Li}
\affiliation{Department of Physics, Tsinghua University, Beijing 100084, China}



\begin{abstract}

We present a detailed analysis of the progenitor and its local environment for the recently discovered type II supernova (SN) 2024ggi at a distance of about 6.7~Mpc, by utilizing the pre-explosion images from the Hubble Space Telescope (HST) and \textit{Spitzer} Space Telescope. The progenitor is identified as a red, bright variable star, with absolute $F814W$-band magnitudes being $-$6.2 mag in 1995 to $-$7.2 mag in 2003, respectively, consistent with that of a normal red supergiant (RSG) star. 
Combining with the historical mid-infrared light curves, a pulsational period of about 379~days can be inferred for the progenitor star. Fitting its spectral energy distribution with stellar spectral models yields the stellar parameters of  temperature, radius and bolometric luminosity as $T_*=3290_{-27}^{+19}$~K, $R_*=887_{-51}^{+60}$~\rsun, and log($L$/\lsun)$=4.92_{-0.04}^{+0.05}$, respectively. The above parameters indicate that the progenitor of SN 2024ggi is consistent with the stellar evolutionary track of a solar-metallicity massive star with an initial mass of $13_{-1}^{+1}$~\msun.
Moreover, our analysis indicates a relatively low mass loss rate (i.e., $< 3\times10^{-6}$~\msun~yr$^{-1}$) for the progenitor compared to that inferred from the flashed spectra and X-ray detection (i.e., $10^{-2}$$-$$ 10$$^{-5}$~\msun~yr$^{-1}$), implying a significant enhancement in mass loss  within a few years prior to the explosion. 

\end{abstract}


\section{Introduction} \label{sec:intro}
Type II Supernovae (SNe II) are thought to result from core-collapse explosions of red supergiants (RSGs), which have initial masses of 8--25~\msun\ \citep{2003ApJ...591..288H}. These stars retain most of their hydrogen envelopes before core collapse, producing supernovae with persistent and prominent hydrogen lines in their spectra. 
With pre-discovery images, mainly thanks to space-based telescopes such as the Hubble Space Telescope (\textit{HST}), progenitors have been identified for dozens of SNe II \citep{2015PASA...32...16S} and more recently for SN~2017eaw 
\citep{2018MNRAS.481.2536K,2019ApJ...875..136V,2019MNRAS.485.1990R}, SN~2022acko \citep{2023MNRAS.524.2186V} and SN~2023ixf \citep{2023ApJ...952L..23K,2023ApJ...952L..30J,2023ApJ...957...64S,2023ApJ...955L..15N,2024SCPMA..6719514X}.
These observations and analysis have confirmed the connections between RSGs and SNe IIP/IIL. 
Some of the progenitors were found to hold thick circumstellar (CS) materials or dust shells which led to suppressed observed flux in optical bands but enhanced one in infrared (IR) bands, e.g., SN~2012aw, SN~2017eaw, and SN~2023ixf. Among them, the dust mass around the progenitor star of SN~2023ixf can even be comparable to that of asymptotic giant branch stars, causing its colors in the IR bands to deviate remarkably from those of ordinary RSGs. 
CS materials locating close to the precursor star can also produce observational spectral features in young SNe~II, typically narrow emission lines in early spectra, the so-called ``flash'' lines \citep{2002ApJ...572..932P,2007ApJ...666.1093Q,2019MNRAS.485.1990R,2023SciBu..68.2548Z}. Moreover, interaction of SN ejecta with CS materials can also result in higher temperature and fast rise in the early light curves. Most interestingly, the serendipitous capture of pre- and after-explosion of SN~2023ixf have witnessed that the resulted shock passed through the dust shell, leading to a quick change of shock radiation from red to blue light \citep{2024Natur.627..754L}. 
It is possible that the close-by CS materials are common for SNe IIP/IIL, but the mass loss rate derived from their spectra or light curves are far higher than that from their RSG progenitors. It remains unknown on how the mass loss of RSGs is enhanced in their final years prior to explosion.

SN~2024ggi is another type II SN discovered within about 7~Mpc after SN~2023ixf over the past decade. It was discovered by ATLAS in the nearby spiral galaxy NGC~3621 ($z=0.002435$) on Apr. 11.14, 2024 \citep{2024TNSAN.100....1S}. The earliest spectrum obtained at about 11 hours after discovery showed prominent narrow emission lines of H, HeI, HeII, CIII, CIV and NIII \citep{2024TNSAN.104....1Z}, and fast evolution of these narrow features was revealed by later high-resolution spectra \citep{2024arXiv240502274P}. 
X-ray emission was also detected by the Follow-up X-ray Telescope (FXT) on board Einstein Probe (EP) \citep{2024ATel16588....1Z}.

Based on the \textit{HST} archival images, \cite{2024TNSAN.100....1S} reported possible detection of a red progenitor of SN~2024ggi. \cite{2024TNSAN.105....1Y} and \cite{2024TNSAN.107....1P} also reported the $griz$-, J- and Ks-band photometry at the SN site, where the progenitor candidate is found to be bright in near-infrared (NIR) bands.

In this letter, we present measurements of \textit{HST} archived pre-explosion images around the site of SN~2024ggi, including the progenitor candidate. We also checked the $Spitzer$/IRAC observations, although they have lower spacial resolutions. With these data, we analyzed the progenitor environment of SN~2024ggi and estimated the properties of its progenitor star. 
A distance modulus to the SN of $\mu=$29.14$\pm$0.06~mag ($D_{\mathrm{L}}=$6.72$\pm$0.19~Mpc) \citep{2013AJ....146...86T} is adopted throughout this paper. 

\section{Observation Data and Reduction} \label{sec:data}

\subsection{\textit{HST} Pre-explosion Images}\label{sec:HST}
We searched the pre-explosion \textit{HST} images from Mikulski Archive for Space Telescopes (MAST)\footnote{\url{http://archive.stsci.edu/}} and the Hubble Legacy Archive (HLA)\footnote{\url{http://hla.stsci.edu/}}, and found publicly available images in various bands, covering the phases from late 1994 to 2003.

\begin{figure*}
    \centering
    \includegraphics[width=0.25\linewidth]{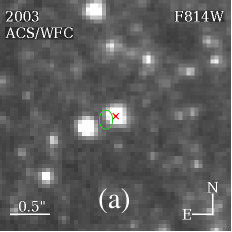}
    \includegraphics[width=0.25\linewidth]{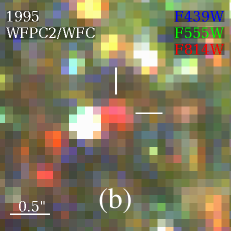}
    \includegraphics[width=0.25\linewidth]{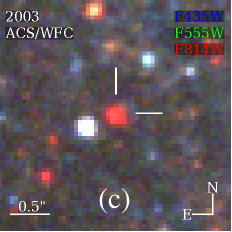}
    \includegraphics[width=0.25\linewidth]{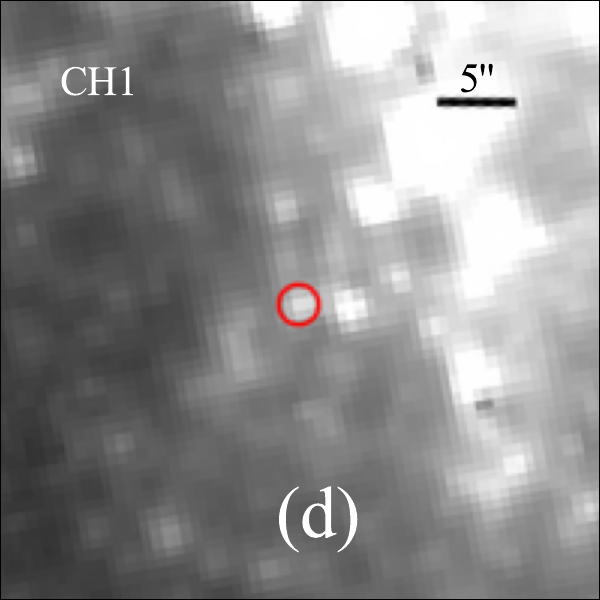}
    \includegraphics[width=0.25\linewidth]{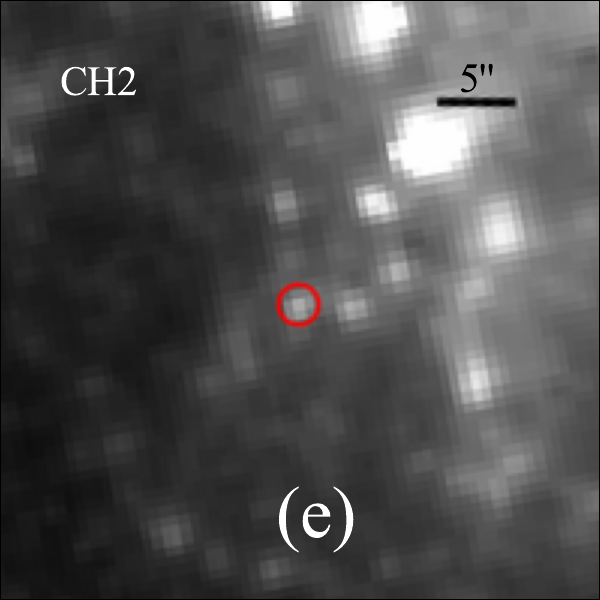}
\caption{(a): The ACS-F814W image centered on the location the progenitor candidate (marked by a red cross). The green ellipse shows the transformed SN position with the size presenting the 1-$\sigma$ error ($\sigma_{\mathrm{RA}}$=0.087\arcsec, $\sigma_{\mathrm{Decl.}}$=0.118\arcsec); (b)$\sim$(c): Pseudo-color images created from the \textit{HST} pre-explosion images obtained in 1995 and 2003. The progenitor candidate is marked by white ticks. 
(d)$\sim$(e): The field of SN~2024ggi on the \textit{Spitzer} CH1 and CH2 images. The red circle in each panel is centered on the SN site with a radius of 1.2\arcsec, which is the aperture of the photometry. North is up and east is to the left.} \label{fig:hstimg}
\end{figure*}

To get accurate positions of \mbox{SN~2024ggi} in the pre-discovery images, we made use of a drizzled ACS/WFC F814W image in 2003 obtained from HLA as a pre-explosion image, and an image combined from 3 unfiltered 3-second images obtained by the Lijiang 2.4-m Telescope (LJT; \citealp{2019RAA....19..149W}) on Apr. 12 2024 as a post-explosion image. We first chose 10 common stars that appeared on the LJT and \textit{HST} images and then got their positions in each image using \texttt{SExtractor} \citep{1996A&AS..117..393B}.
A second-order polynomial geometric transformation function is applied using the IRAF\footnote{IRAF is distributed by the National Optical Astronomy Observatories, which were operated by the Association of Universities for Research in Astronomy, Inc., under cooperative agreement with the National Science Foundation (NSF).} \texttt{geomap} task to convert their coordinates in the post-explosion image to those in the pre-explosion images.
Based on the \texttt{IRAF} \texttt{geoxytran} task, this established the transformation relationship between the coordinates of SN~2024ggi in the post-explosion image and those in the \textit{HST} pre-explosion images.
The uncertainties of the transformed coordinates are a combination of the uncertainties of the SN position and the geometric transformation. The identified progenitor candidate in the pre-explosion images is shown in Fig.~\ref{fig:hstimg}(a). The point spread function (PSF) of the progenitor candidate in the ACS-F814W image has a full width at half maximum (FWHM) of $\sim$8.5 pixels ($\sim$0.43\arcsec).
The locations of the progenitor in other band images are obtained by either visually matching the images to the F814W image or transforming from the F814W image using similar coordinates transform procedures. In Fig.~\ref{fig:hstimg}(b)$\sim$(c) we show the pseudo-color images created from the \textit{HST} pre-explosion images obtained in 1995 and 2003.

DOLPHOT\footnote{\url{http://americano.dolphinsim.com/dolphot/}} 2.0 was used to get photometry of the progenitor on the pre-explosion images. The photometry is performed on the bias- and flat-corrected C0M FITS images from the WFPC2/WFC instrument, and bias-, flat-, CTE-corrected FLC FITS images from the ACS/WFC and ACS/WFC instruments, all obtained from the \mbox{MAST} archive.
Choosing the F814W image as a reference image, DOLPHOT is run simultaneously on multiple-band images taken on the same day.
The progenitor is not detected in the WFC3/UVIS images in all bands and the WFPC2/WFC images in F336W, F439W and F658N bands.
Magnitudes and their uncertainties of the progenitor candidate are extracted from the output of DOLPHOT. The photometric results are listed in Tab.~\ref{tab:hst-obs}.

Multiple observations were conducted by WFPC2 in F555W and F814W bands during the period from Dec. 27 1994 to Feb. 12 1995. And observations on Feb. 3 2003 repeated several times in ACS F435W, F555W and F814W bands. For these images, we also performed photometry on the combined images of each band. These results are given in Tab.~\ref{tab:hst-obs-comb}. 

As shown in Tab.~\ref{tab:hst-obs-comb}, from 1995 to 2003, the progenitor brightened by 0.99$\pm$0.02 mag and 0.80$\pm$0.39 mag in F814W and F555W bands, respectively. 
As discussed in Appendix~\ref{appendix:field-variable}, the variability in F814W band is not due to any artificial effect but intrinsic to pulsational RSG progenitor, while the result shown in F555W band is less credible.

A reddening of $E(B-V)$ = 0.113$\pm$0.023 mag can be inferred for the host galaxy from the Na~I~D lines in the mid-resolution spectra of SN~2024ggi \citep[][in Prep.]{ZhangJJ2024ggi}. Including the Galactic reddening of $E(\mathrm{B-V})=$0.072~mag \citep{2011ApJ...737..103S}, the total reddening to SN~2024ggi is given as $E(B-V)=0.185\pm0.023$~mag.
The total extinction is consistent with that derived by \cite{2024arXiv240502274P}, which is $E(B-V)$ = 0.16$\pm$0.02.
Applying the distance to NGC~3621 and the above total reddening, we get absolute magnitudes in each bands.
The progenitor is very red with $M_{\mathrm{F555W}} = -1.71\pm0.67$, $M_{\mathrm{F814W}} = -6.21\pm0.08$ mag, F555W$-$F814W$=4.50\pm0.66$~mag in 1995, and $M_{\mathrm{F555W}} = -2.53\pm0.26$, $M_{\mathrm{F814W}} = -7.20\pm0.07$~mag, F555W$-$F814W$=4.67\pm0.25$~mag in 2003, respectively.
Such a large color index immediately reminds us of the most striking SN II in 2023: SN~2023ixf, whose progenitor was found to have F555W$-$F814W$=4.28\pm1.23$~mag, but with a much fainter $M_{\mathrm{F814W}}$ of $-4.84\pm0.05$~mag.
Therefore, the severely suppressed flux in blue bands is also likely due to that the surrounding dust obscuring the progenitor.

Close inspections of the progenitor environment on the HST images reveals that there are several bright stars near the progenitor, among which the closest one is about 0.39\arcsec\ away. Given a distance of 6.7~Mpc to NGC 3621, this corresponds to a spacial distance of $\sim$12.7~pc. However, this neighbor star appears much bluer (with a blackbody temperature of over 10,000~K) than the progenitor candidate. The Legacy Survey pre-explosion $griz$ images \citep{2024TNSAN.105....1Y} have a resolution of 0.1\arcsec/pixel, but photometry (with $g-r\approx0$~mag) at the SN site is mainly attributed to the neighbor star rather than the progenitor candidate.

\subsection{\textit{Spitzer}-IRAC observations and data reduction}\label{sec:spitzer}

The SN~2024ggi field in NGC 3621 was observed with the \textit{Spitzer} Infrared Array Camera (IRAC) in mid-infrared (MIR) bands before its explosion. The observations were monitored by several programs covering from 2004 to 2019, by PI R.Kennicutt (program ID: 159) and M. Kasliwal (program IDs: 10136, 11063, 13053, 14089). The level 2 post-BCD (Basic Calibrated Data) images were obtained from the \textit{Spitzer} Heritage Archive (SHA)\footnote{\url{http://irsa.ipac.caltech.edu/applications/Spitzer/SHA/}}, which were reduced by the \textit{Spitzer} pipeline and resampled onto 0.6\arcsec/pixel. A point-like source can be detected in $CH1$ ($3.6\mu$m) and $CH2$ ($4.5\mu$m) images near the site of SN~2024ggi with a $2\sigma$ threshold (Fig.~\ref{fig:hstimg}~(d)$\sim$(e)).

Aperture photometry was performed on the pre-explosion images of the SN field with an aperture radius of 2 pixels (1.2\arcsec). Aperture corrections were applied following the IRAC Data Handbook. The level 2 post-BCD images have been calibrated in an absolute surface-brightness unit of MJy/sr, which can be transformed into units of $\mu$Jy/pixel$^2$ by a conversion factor of 8.4616 for the angular resolution of the IRAC images 0.6\arcsec/pixel.
The measured fluxes in CH1 and CH2 bands are listed in Tab.~\ref{tab:spitzer}. 

However, we caution that the PSF of the progenitor candidate has an FWHM of about 2 pixels, i.e. $\sim$1.2\arcsec, which is about 5.6 times that of the star in the \textit{HST} images. In this radius range, several bright stars are found on the \textit{HST} images, and resolved photometry of the progenitor on the \textit{Spitzer} images is not possible (as shown in Fig.~\ref{fig:HST-stars-img} and the magnitudes of the bright stars are tabulated in Tab.~\ref{tab:HST-fieldstars}). 
On the other hand, \cite{2024TNSAN.107....1P} reported photometric results of $m_J=19.74\pm0.17$~mag and $m_{\mathrm{Ks}}=18.44\pm0.30$~mag for the progenitor candidate using the pre-explosion images of VISTA Hemisphere Survey \citep[VHS,][]{2013Msngr.154...35M} with an aperture radius of 1\arcsec. The VHS images have better spacial resolution than the \textit{Spitzer} ones.

\section{Data analysis} \label{sec:analysis}

\subsection{Infrared emission of the progenitor star}\label{sec:prog-ir-flux}
As noted in Sec.~\ref{sec:spitzer}, measurements of the flux in the pre-explosion \textit{Spitzer} images at the SN site are probably contributed by several stars including the progenitor of SN~2024ggi. 
Inspecting the colors and magnitudes of stars within 1.2\arcsec\ around the progenitor candidate, we found that several stars may contribute to the flux in the IR bands. As shown in Fig.~\ref{fig:HST-star-CM}, although the progenitor is the brightest and reddest star, there are at least 2 other bright stars with comparable red colors. Nonetheless, it is still possible to estimates the contribution of the progenitor in IR bands. As discussed in Appendix~\ref{appendix:field-stars}, the measured NIR and MIR fluxes mainly come from the contribution of the progenitor. The nearby stars contribute only $\lesssim$12\% and $\lesssim$30\% to the total fluxes in NIR and MIR bands, respectively.

Phase-averaged flux in CH1 and CH2 bands are $f_{\mathrm{CH1}}=10.36\pm0.96$~$\mu$Jy, $f_{\mathrm{CH2}}=9.25\pm0.56$~$\mu$Jy, giving $m_{\mathrm{CH1}}=18.56\pm0.10$~mag, $m_{\mathrm{CH1}}=18.21\pm0.07$~mag. The corresponding absolute magnitudes are $M_{\mathrm{CH1}}=-10.62\pm0.12$~mag and $M_{\mathrm{CH2}}=-10.96\pm0.09$~mag, respectively.
Then, after subtraction of the field stars, the $M_{\mathrm{CH1}}$ and $M_{\mathrm{CH2}}$ values of the progenitor are $-10.43\pm0.21$~mag and $-10.77\pm0.12$~mag, respectively.
Using the results from \cite{2024TNSAN.107....1P}, we obtain $M_{\mathrm{J}}=-9.47\pm0.20$~mag and $M_{\mathrm{Ks}}=-10.72\pm0.32$~mag for the progenitor star.
These absolute magnitudes and colors are well within the range of RSGs \cite[e.g.][]{2020A&A...639A.116Y}.

\subsection{Periodic variability in Mid-infrared and optical bands}\label{sec:period}
\begin{figure}
    \centering
    \includegraphics[width=1.0\linewidth]{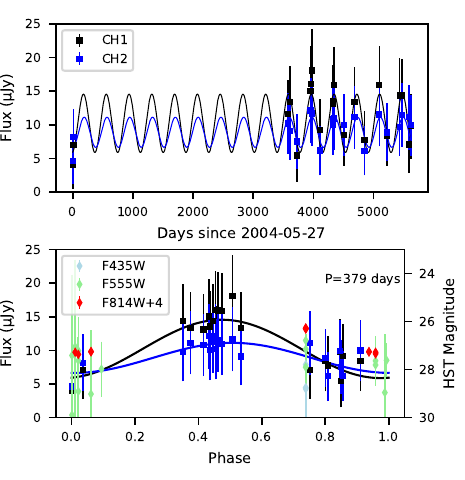}
    \caption{Periodic analysis of the \textit{Spitzer} CH1 and CH2 light curves. \textit{Upper}: Light curves in CH1 and CH2 bands, the fitted periodic light curves with $P=379$~days are plotted as solid lines; \textit{Lower}: The phase-folded MIR light curves overplotted with the \textit{HST} magnitudes of the progenitor.}
    \label{fig:lc-period}
\end{figure}

As discussed in Sec.~\ref{sec:prog-ir-flux}, the MIR flux measured from \textit{Spitzer} CH1 and CH2 images is mainly contributed by the progenitor of SN~2024ggi.
The \textit{Spitzer} light curves show prominent temporal variation (top panel of Fig.~\ref{fig:lc-period}).
We searched for periodic variability by applying the Lomb-Scargle method \citep{1976Ap&SS..39..447L,1982ApJ...263..835S} with the \texttt{python} routine \texttt{LombScargleMultiband} in the package \texttt{astropy.timeseries} \citep{2012cidu.conf...47V,2015ApJ...812...18V} on the MIR light curves. A period of 378.5$\pm$29.4~days (log$P\sim2.58$) can be found by fitting the light curves in CH1 and CH2 bands simultaneously. 
Amplitudes of the light curves are $\sim$0.49~mag and $\sim$0.28~mag in CH1 and CH2 bands, respectively. 
This variability is most likely due to radial pulsation of the progenitor star.

If we assume that the variation in MIR bands is attributed to the pulsation of the progenitor of SN~2024ggi, with empirical period-luminosity ($P-L$) relations of RSGs, we are able to estimate its absolute magnitudes in IR bands. However, as noted by \cite{2024SCPMA..6719514X}, dusty RSGs may deviate from the common $P-L$ relation of RSG groups at wavelengths shorter than CH1 due to possible dust absorption. Absolute magnitude in CH1 band is derived as $-10.14$~mag or $-10.38$~mag if the $P-L$ relation of RSGs in M33 or M31 is applied \citep[see Tab.~6 and 7 of][]{2019ApJS..241...35R}. This value is very close to that derived in Sec.~\ref{sec:prog-ir-flux}. We also checked the magnitudes in other bands, and found that they all obey the $P-L$ relation of RSGs. This suggests that the progenitor of SN~2024ggi be a normal RSG in IR bands.

Temporal variation is also seen in \textit{HST} optical bands (see Tab.~\ref{tab:hst-obs}). The \textit{HST} phases and magnitudes are plotted in Fig.~\ref{fig:lc-period}. As shown in Fig.~\ref{fig:lc-period}, changes in optical bands are synchronized with those seen in the IR bands.
The epoch of observations in 2003 corresponds to a phase near 0.75, and the epochs in 1995 correspond to phases near the minimum of periodic variation. Therefore, the amplitudes in optical bands are roughly the differences of the magnitudes between 1995 and 2003, i.e., 0.80$\pm$0.39~mag and 0.99$\pm$0.03~mag in F555W and F814W band, respectively.
The variation amplitudes decrease with increasing wavelengths, consistent with other long period variables \citep[e.g.][]{2015MNRAS.447.3909R,2018A&A...616A.175Y}. 

\subsection{Local environment of the progenitor star} \label{sec:host-env}
The host galaxy of SN~2024ggi is a face-on spiral galaxy with an extended disk \citep{Koribalski_2004}. The site of SN explosion is 110\arcsec\ ($\sim$3.6~kpc) away from the galactic center. According to the metallicity measurement of NGC~3621, the local metallicity of SN~2024ggi has a range of [Fe/H]=$-$0.18$\sim$0.16 with different methods adopted \citep{2012ApJ...750..122B}.

\begin{figure}
    \centering
    \includegraphics[width=0.9\linewidth]{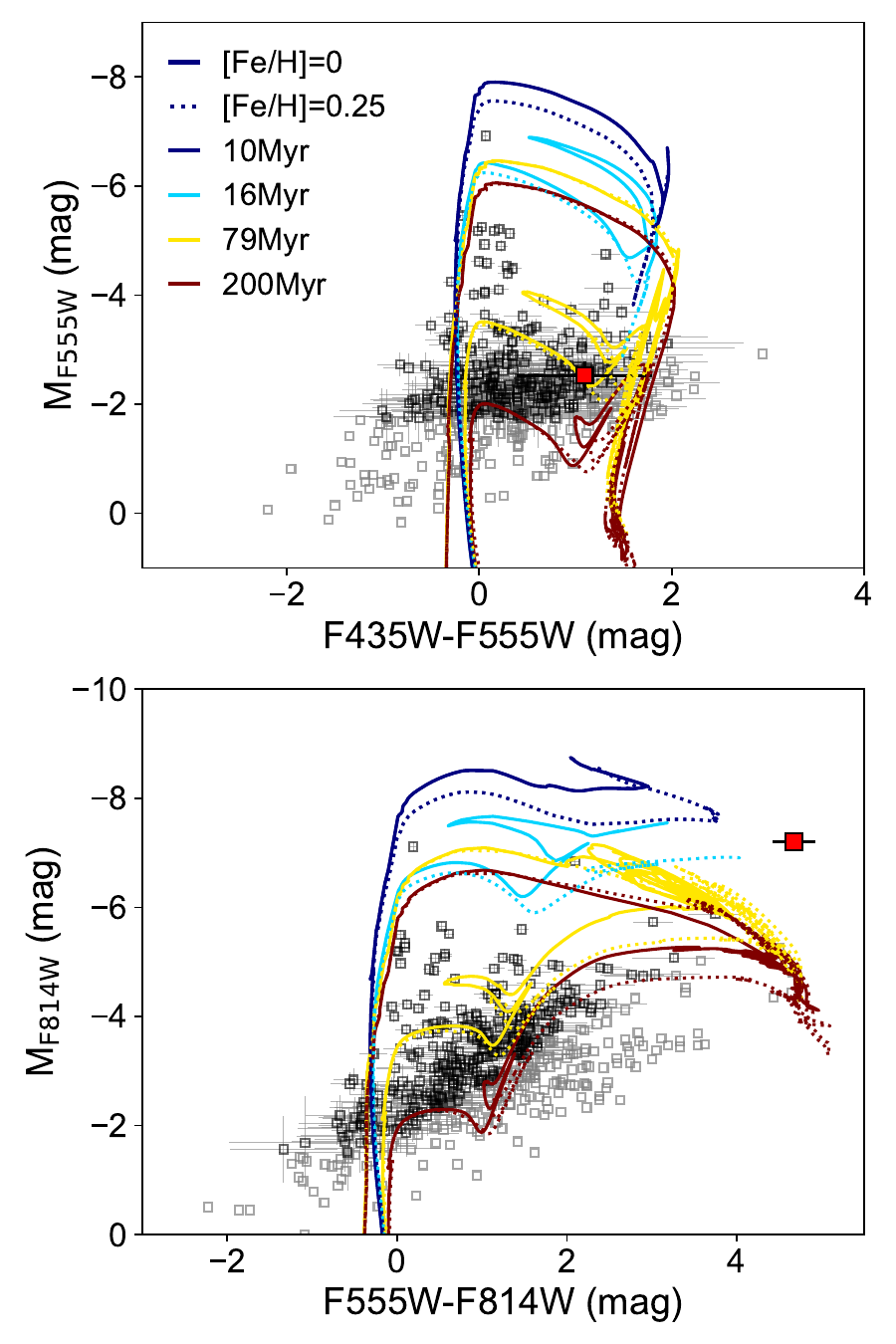}
    \caption{The extinction-corrected color and absolute magnitude diagram of stars within 3\arcsec\ (100~pc) around the progenitor. The progenitor is marked by a red filled square in each panel. \textit{Left}: F435W$-$F555W \textit{vs.} $M_\mathrm{F555W}$. Faint stars with $m_{\mathrm{F435W}}>$29~mag or $m_{\mathrm{F555W}}>$28~mag are plotted as gray squares; \textit{Right}: F555W$-$F814W \textit{vs.} $M_{\mathrm{F814W}}$. Faint stars with $m_{\mathrm{F555W}}>$28~mag or $m_{\mathrm{F814W}}>$28~mag are plotted as gray squares. For comparison, MIST isochrones with [Fe/H]=0 (solid lines) and [Fe/H]=0.25 (dotted lines) at different ages are overplotted. }
    \label{fig:HST-SP}
\end{figure}

When looking into the stellar population within 100~pc around the SN site in Fig.~\ref{fig:HST-SP}, we found that the stellar population around the SN site seems to cover a large age range. Compared with MESA Isochrones and Stellar Tracks \citep[MIST;][]{2016ApJ...823..102C,2016ApJS..222....8D}, the bluest stars belong to a young stellar population of $t=$10~Myr, but most stars are red and faint and belong to a much older population of $t=$200~Myr.
Such multiple age components were also found in the surrounding stellar populations of some SNe IIP \citep{2017MNRAS.469.2202M}.
Among the red stars, the progenitor is the reddest and distinct from any populations. Higher metallicity may solve this problem. With [Fe/H]=0.25, the closest population to the progenitor has an age of 16~Myr, which is approximately the life time of the progenitor of a type II SN. 
Thus, we assume solar metallicity in our analysis below.

\subsection{Properties of the progenitor star}\label{sec:sed-fit}
\begin{figure}
    \centering
    \includegraphics[width=1.0\linewidth]{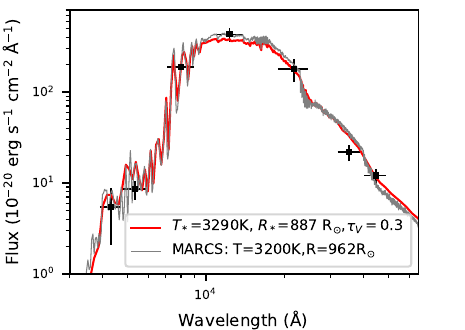}
     \includegraphics[width=1.0\linewidth]{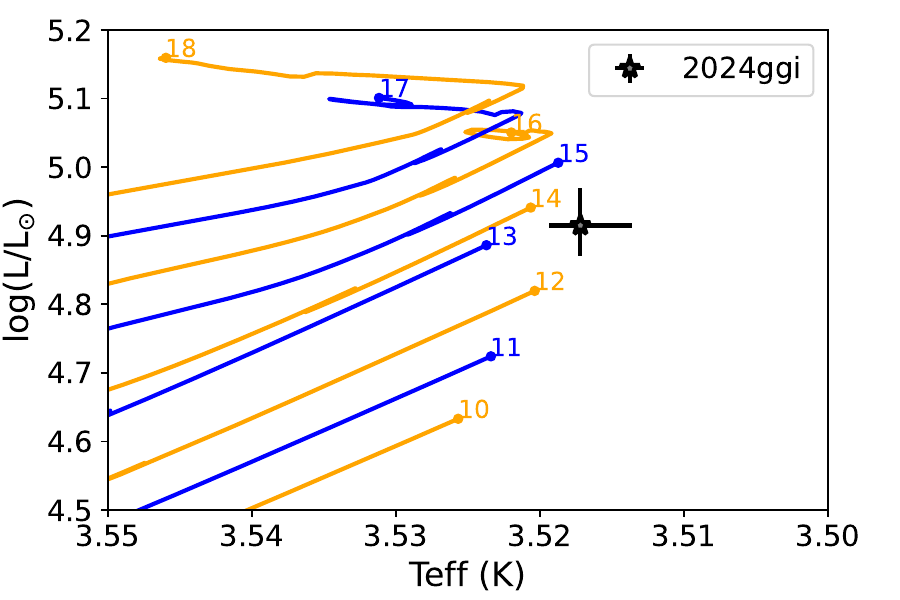}
    \caption{\textit{Upper}: Spectral energy distribution of the progenitor of SN~2024ggi (black squares) and the best-fit DUSTY+MARCS model with log~g = $-$0.5, $R_{\mathrm{out}}/R_{\mathrm{in}}$ = 2.0. A dust-free MARCS stellar model with [Fe/H]$=0$, $T_{\mathrm{eff}}=3200$~K is also plotted as a grey line. \textit{Lower}: Location of the progenitor of SN~2024ggi on the HRD. Overplotted are the MIST stellar evolutionary tracks of [Fe/H]$=0$, $v/v_{\mathrm{crit}}=0.4$. The filled dots marked by the initial masses (in solar mass) represent the ending points of the MIST tracks.}
    \label{fig:SED-fit}
\end{figure}

With the measured photometric data of the progenitor through optical to MIR bands, we are able to fit the SED of the progenitor.
First we check dust-free stellar spectral models. 
The MARCS spectra models\footnote{\url{https://marcs.oreme.org/}} \citep{2008A&A...486..951G,2017A&A...601A..10V} are compared with the SED of the progenitor. As shown in Fig.~\ref{fig:SED-fit}, the SED of the progenitor can be well fit by the spectral model with a relatively low temperature of $T_{\mathrm{eff}}=3200$~K. The corresponding stellar radius is about 962~\rsun. Such a temperature, however, is relatively low among other SN IIP progenitors \citep{2015PASA...32...16S}.

\begin{table*}
\centering
\caption{MCMC fitting parameters for the SED of the progenitor of SN~2024ggi with DUSTY+MARCS models. 
The mass loss rate is calculated by assuming a wind velocity of 50~km~s$^{-1}$. The lower and upper limits are given as the values at 16\%, 84\% of the posterior probability distribution of the MCMC sampling. The last column presents the least $\chi^2$ of the best-fit model. \label{tab:fit-res}}
\begin{tabular}{ccccccccccc}
\toprule
[Fe/H] &log~g &$R_{\mathrm{out}}/R_{\mathrm{in}}$ &$T_*$ &$\tau_V$ &log($L$/\lsun) &$R_d$ &$R_*$ &$\dot{M}$ & $M_{\mathrm{w}}$ &$\chi^2$\\
       &      &                                   &(K)   &         &    &(10$^{16}$~cm) &(\rsun) &(10$^{-6}$~\msun~yr$^{-1}$) &(10$^{-5}$~\msun) & \\
\hline
0.00 &$-$0.5 & 2 &3290$_{-27}^{+16}$ &0.26$_{-0.19}^{+0.40}$ &4.92$_{-0.02}^{+0.03}$ &1.00$_{-0.97}^{+30.53}$ &887$_{-26}^{+35}$ &0.83$_{-0.63}^{+1.63}$ &7.18$_{-6.97}^{+564.20}$ &2.816\\
0.00 &$-$0.5 &10 &3289$_{-28}^{+19}$ &0.22$_{-0.16}^{+0.32}$ &4.92$_{-0.02}^{+0.03}$ &1.45$_{-1.40}^{+35.93}$ &892$_{-28}^{+35}$ &1.01$_{-0.76}^{+1.88}$ &92.47$_{-91.06}^{+4682.00}$ &2.815\\
0.00 &0.0 & 2 &3276$_{-28}^{+34}$ &0.55$_{-0.43}^{+0.76}$ &4.95$_{-0.03}^{+0.03}$ &2.52$_{-2.48}^{+44.10}$ &926$_{-33}^{+38}$ &1.96$_{-1.55}^{+3.36}$ &63.72$_{-63.31}^{+2278.00}$ &4.255\\
0.00 &0.0 &10 &3276$_{-30}^{+35}$ &0.40$_{-0.32}^{+0.56}$ &4.95$_{-0.03}^{+0.03}$ &2.59$_{-2.51}^{+44.13}$ &926$_{-33}^{+39}$ &2.05$_{-1.70}^{+3.71}$ &376.90$_{-374.60}^{+10840.00}$ &4.322\\
\hline
\end{tabular}
\end{table*}
Similar to SN~2023ixf, the flash lines in the early spectra of SN~2024ggi reveal possible presence of dust shell surrounding its progenitor star. We then fit the SED with stellar spectral models with dust.
The fitting method is the same as in \cite{2024SCPMA..6719514X} (see also \citealp{2018MNRAS.481.2536K,2021ApJ...912..112W}).
First, we use DUSTY \citep{1997MNRAS.287..799I} to calculate the output flux of a star with spectral models from the MARCS spectral library with [Fe/H]=0.
We assume that the dust is 100\% graphite and a gas to dust ratio $\psi=200$.
The model parameters of DUSTY have optical depth of $\tau_V=10^{-4}\sim10$. The ratio of the outer and inner radius of the dust shell $R_{\mathrm{out}}/R_{\mathrm{in}}$ = 2, 10. For RSGs, log~g = $-$0.5 and 0.0 are adopted.
The fitting results are reported in Tab.~\ref{tab:fit-res}, and the best-fit model (with least $\chi^2$) is shown in the upper panel of Fig.~\ref{fig:SED-fit}.

As shown in Tab.~\ref{tab:fit-res}, models with log~g = $-$0.5 have the minimum $\chi^2$. 
According to the fitting, the progenitor has a radius $R_*=887_{-51}^{+60}$~\rsun, $T_*=3290_{-27}^{+19}$~K, and log($L$/\lsun)$=4.92_{-0.04}^{+0.05}$.\footnote{Uncertainty of the distance is included in the uncertainties of $R_*$ and log($L$/\lsun). }
Comparing the location of the progenitor with MIST single star evolutionary tracks on the Hertzsprung–Russell diagram (HRD), the progenitor star is estimated to have an initial mass of $13_{-1}^{+1}$~\msun\ (see the lower panel of Fig.~\ref{fig:SED-fit}).

The optical depth $\tau_V\approx0.3$ implies that the dust shell is optically thin. 
Even though the fitted model can match the observation well, we note that the posterior probability distribution of $\tau_V$ is maximized at the lower limit of the model parameter grid. This indicates that dust has very limited influence on the SED of the progenitor.
The mass loss rate yielded for the progenitor has an upper limit of 3$\times10^{-6}$~\msun~yr$^{-1}$.
This rate is much less than that derived from the SN flash lines \citep[$\sim10^{-2}$~\msun~yr$^{-1}$;][]{2024arXiv240419006J} or X-ray light curves \citep[$\sim10^{-5}$~\msun~yr$^{-1}$;][In Prep.]{24ggiLCEP}. Similar to SN~2023ixf, this implies that the progenitor of SN~2024ggi also had an enhanced mass loss prior to the explosion.

\section{Conclusion} \label{sec:conclusion}

SN~2024ggi is another nearby H-rich supernova which presents strong flash lines in its early spectra after SN~2023ixf. 
The SN is located in the inner disk of the spiral galaxy NGC~3621.
We studied the progenitor candidate of SN~2024ggi with pre-explosion \textit{HST} images in optical bands and \textit{Spitzer} images in mid-infrared (MIR) bands. 

The progenitor star is clearly resolved on the pre-explosion \textit{HST} images obtained over 20 years before the explosion. The progenitor candidate is the reddest star around the SN site and the brightest among the red stars. 
A bright point-like source can be seen at the SN site on the \textit{Spitzer}/IRAC images in CH1 and CH2 bands. Although the progenitor can not be resolved from the \textit{Spitzer}/IRAC images, over 70\% of the flux measured by aperture photometry is contributed by the progenitor star. 
The MIR fluxes show periodic variation with a period of 379~days, amplitudes of 0.49~mag and 0.28~mag in CH1 and CH2 bands, respectively.
Similar temporal variation can be found in \textit{HST} light curves. Applying the period found from MIR light curves, variation amplitudes in F555W and F814W bands are estimated to be $\sim$0.8~mag and $\sim$1~mag, respectively. 
All these observational properties agree well with red supergiants (RSGs).

Further study of the progenitor star with stellar spectral models resulted in stellar parameters of $T_*\approx3290$~K, $R_*\approx887$~\rsun, and log($L$/\lsun)$\approx4.92$. The possible dust shell surrounding the progenitor is found to be very thin ($\tau_V\sim0.3$), implying a much thinner circumstellar shell than that of SN~2023ixf.
The derived mass loss rate is much lower than that deduced from the early observations of the SN, indicating that the progenitor might have experienced a significantly enhanced mass loss when approaching its death.
The location of the progenitor on the Hertzsprung–Russell diagram is consistent with stellar evolutionary tracks with an initial mass of $13_{-1}^{+1}$~\msun\ at solar metallicity.


\begin{acknowledgments}
This work is supported by the National Natural Science Foundation of China (NSFC grants 12288102, 12033003, 11633002, and 12303047) and the Tencent Xplorer Prize. L.W. is sponsored (in part) by the Chinese Academy of Sciences (CAS), through a grant to the CAS South America Center for Astronomy (CASSACA) in Santiago, Chile. J.Z. is supported by the National Key R\&D Program of China with No. 2021YFA1600404, the National Natural Science Foundation of China (12173082), the Yunnan Province Foundation (202201AT070069), the Top-notch Young Talents Program of Yunnan Province, the Light of West China Program provided by the Chinese Academy of Sciences, and the International Centre of Supernovae, Yunnan Key Laboratory (No. 202302AN360001).
We acknowledge the support of the staff of the LJT. Funding for the LJT has been provided by the CAS and the People's Government of Yunnan Province. The LJT is jointly operated and administrated by YNAO and the Center for Astronomical Mega-Science, CAS.
\end{acknowledgments}

%

\vspace{5mm}
\facilities{\textit{HST} (WFPC2, ACS, WFC3), \textit{Spitzer}/IRAC \citep{spitzer}, VISTA-Vircam, LJT. }


\software{astropy \citep{2013A&A...558A..33A,2018AJ....156..123A}, 
          emcee \citep{2013PASP..125..306F},
          SExtractor \citep{1996A&AS..117..393B},
          IRAF \citep{1986SPIE..627..733T,1993ASPC...52..173T},
          DOLPHOT (\url{http://americano.dolphinsim.com/dolphot/}),
          DUSTY version 2.0 \citep{1997MNRAS.287..799I}.
          }



\appendix

\section{Photometric results of \textit{HST} and \textit{Spitzer}/IRAC images} \label{appendix:phot-tab}
Photometric results of the pre-explosion \textit{HST} images of the progenitor star of SN~2024ggi are presented in Tab.~\ref{tab:hst-obs} and Tab.~\ref{tab:hst-obs-comb}. The results on the \textit{Spitzer} images of the progenitor star are presented in Tab.~\ref{tab:spitzer}. 
\begin{table}
	\footnotesize
	\caption{Photometric results of the pre-explosion \textit{HST} images at the site of \mbox{SN~2024ggi}. All magnitudes are in the Vega system.\label{tab:hst-obs}}
	\tabcolsep 4pt 
	\centering
	\begin{tabular}{ccccc}
	\toprule
	Obs. date &Instrument &Filter  &mag/limit &1-$\sigma$ error\\
    \hline
    1994-12-27 &WFPC2/WFC &F814W &23.264 &0.050\\
    1994-12-27 &WFPC2/WFC &F814W &23.251 &0.049\\
    1995-01-04 &WFPC2/WFC &F555W &27.807 &0.865\\
    1995-01-04 &WFPC2/WFC &F555W &27.635 &1.026\\
    1995-01-04 &WFPC2/WFC &F814W &23.276 &0.050\\
    1995-01-04 &WFPC2/WFC &F814W &23.342 &0.055\\
    1995-01-15 &WFPC2/WFC &F555W &28.957 &2.429\\
    1995-01-17 &WFPC2/WFC &F555W &27.612 &0.717\\
    1995-01-20 &WFPC2/WFC &F555W &27.411 &0.575\\
    1995-01-20 &WFPC2/WFC &F555W &29.894 &5.840\\
    1995-01-24 &WFPC2/WFC &F555W &30.127 &6.698\\
    1995-01-24 &WFPC2/WFC &F555W &27.050 &0.437\\
    1995-01-24 &WFPC2/WFC &F814W &23.275 &0.049\\
    1995-01-28 &WFPC2/WFC &F555W &28.907 &2.263\\
    1995-01-28 &WFPC2/WFC &F555W &27.627 &0.696\\
    1995-01-28 &WFPC2/WFC &F555W &27.217 &1.417\\
    1995-01-28 &WFPC2/WFC &F814W &23.367 &0.139\\
    1995-02-12 &WFPC2/WFC &F555W &29.019 &2.649\\
    1995-02-12 &WFPC2/WFC &F814W &23.252 &0.047\\
    1995-02-12 &WFPC2/WFC &F814W &23.253 &0.047\\
    1995-02-25 &WFPC2/WFC &F555W &27.982 &1.150\\
    2003-02-03 &ACS/WFC &F435W &28.749 &1.367\\
    2003-02-03 &ACS/WFC &F555W &27.899 &0.706\\
    2003-02-03 &ACS/WFC &F814W &22.275 &0.017\\
    2003-02-03 &ACS/WFC &F435W &27.793 &0.578\\
    2003-02-03 &ACS/WFC &F555W &26.785 &0.278\\
    2003-02-03 &ACS/WFC &F814W &22.309 &0.017\\
    2003-02-03 &ACS/WFC &F435W &28.802 &1.332\\
    2003-02-03 &ACS/WFC &F555W &27.149 &0.368\\
    2003-02-03 &ACS/WFC &F814W &22.272 &0.017\\
    2003-02-03 &ACS/WFC &F814W &22.289 &0.017\\
    \hline
    \end{tabular}
\end{table}
\begin{table}
	\footnotesize
	\caption{Photometric results of the progenitor candidate on the combined pre-explosion \textit{HST} images. All magnitudes are in the Vega system. Detection limits are given at 5-$\sigma$.\label{tab:hst-obs-comb}}
	\tabcolsep 2pt 
	\centering
    \begin{tabular}{ccccc}
    \toprule
	Obs. date range &Instrument &Filter  &mag/limit &1-$\sigma$ error\\
    \hline
    1994-12-27/1995-02-12 & WFPC2/WFC &F814W &23.274 &0.019\\
	1995-01-04/1995-02-25 & WFPC2/WFC &F555W &28.013 &0.302\\
    2003-02-03/2003-02-03 & ACS/WFC &F435W &28.464 &0.661\\
    2003-02-03/2003-02-03 & ACS/WFC &F555W &27.209 &0.245\\
    2003-02-03/2003-02-03 & ACS/WFC &F814W &22.286 &0.008\\
    1995-01-04/1995-01-04 & WFPC2/WFC &F336W &$>$24.7&\\
    1995-01-04/1995-01-24 & WFPC2/WFC &F439W &$>$26.0&\\
    2009-01-17/2009-01-17 & WFPC2/WFC &F658N &$>$22.8&\\
    2019-12-13/2019-12-13 & WFC3/UVIS &F336W &$>$26.3&\\
    2019-12-13/2019-12-13 & WFC3/UVIS &F275W &$>$25.5&\\
    2023-05-19/2023-05-19 & WFC3/UVIS &F300X &$>$26.3&\\
    \hline
\end{tabular}
\end{table}

\begin{table}
\centering
\footnotesize
\caption{The \textit{CH1}- and \textit{CH2}-band aperture photometry on the pre-explosion \textit{Spitzer}/IRAC images at the site of SN~2024ggi. An aperture radius of 1.2\arcsec\ is adopted in the measurement.}\label{tab:spitzer}
\begin{tabular}{ccccc}
\toprule
MJD    &CH1 flux    &$\sigma$  &CH2 flux  & $\sigma$ \\
(days) &($\mu$Jy)   &($\mu$Jy)     &($\mu$Jy) &($\mu$Jy) \\
\hline
53166.08 	&7.04 	&4.27 	&8.17 	&4.21 \\
53152.49 	&3.98 	&3.54 	&4.62 	&3.39 \\
56889.94 	&5.43 	&3.89 	&7.52 	&4.07 \\
56769.85 	&13.36 	&5.36 	&9.10 	&4.31 \\
56740.32 	&11.61 	&5.03 	&10.19 	&4.50 \\
57111.58 	&15.04 	&5.53 	&12.11 	&4.84 \\
57118.96 	&16.03 	&5.72 	&12.09 	&4.83 \\
57139.42 	&18.08 	&6.13 	&11.66 	&4.77 \\
57271.89 	&9.21 	&4.64 	&6.17 	&3.71 \\
57484.19 	&13.10 	&5.23 	&10.79 	&4.61 \\
57491.65 	&13.58 	&5.30 	&10.03 	&4.46 \\
57506.37 	&15.88 	&5.69 	&10.95 	&4.63 \\
57672.07 	&8.44 	&4.51 	&9.97 	&4.56 \\
57847.87 	&13.38 	&5.29 	&11.14 	&4.70 \\
58012.19 	&7.76 	&4.31 	&6.15 	&3.68 \\
58260.46 	&15.94 	&5.80 	&11.54 	&4.73 \\
58389.03 	&8.42 	&4.52 	&8.87 	&4.31 \\
58598.27 	&14.37 	&5.50 	&9.72 	&4.34 \\
58633.65 	&14.36 	&5.50 	&11.42 	&4.72 \\
58749.71 	&7.05 	&4.26 	&11.11 	&4.81 \\
58785.52 	&9.81 	&4.83 	&9.93 	&4.54 \\
\hline
\end{tabular}
\end{table}

\section{Variability of the progenitor and field stars in \textit{HST} bands}\label{appendix:field-variable}
\begin{figure}
    \centering
    \includegraphics[width=0.8\linewidth]{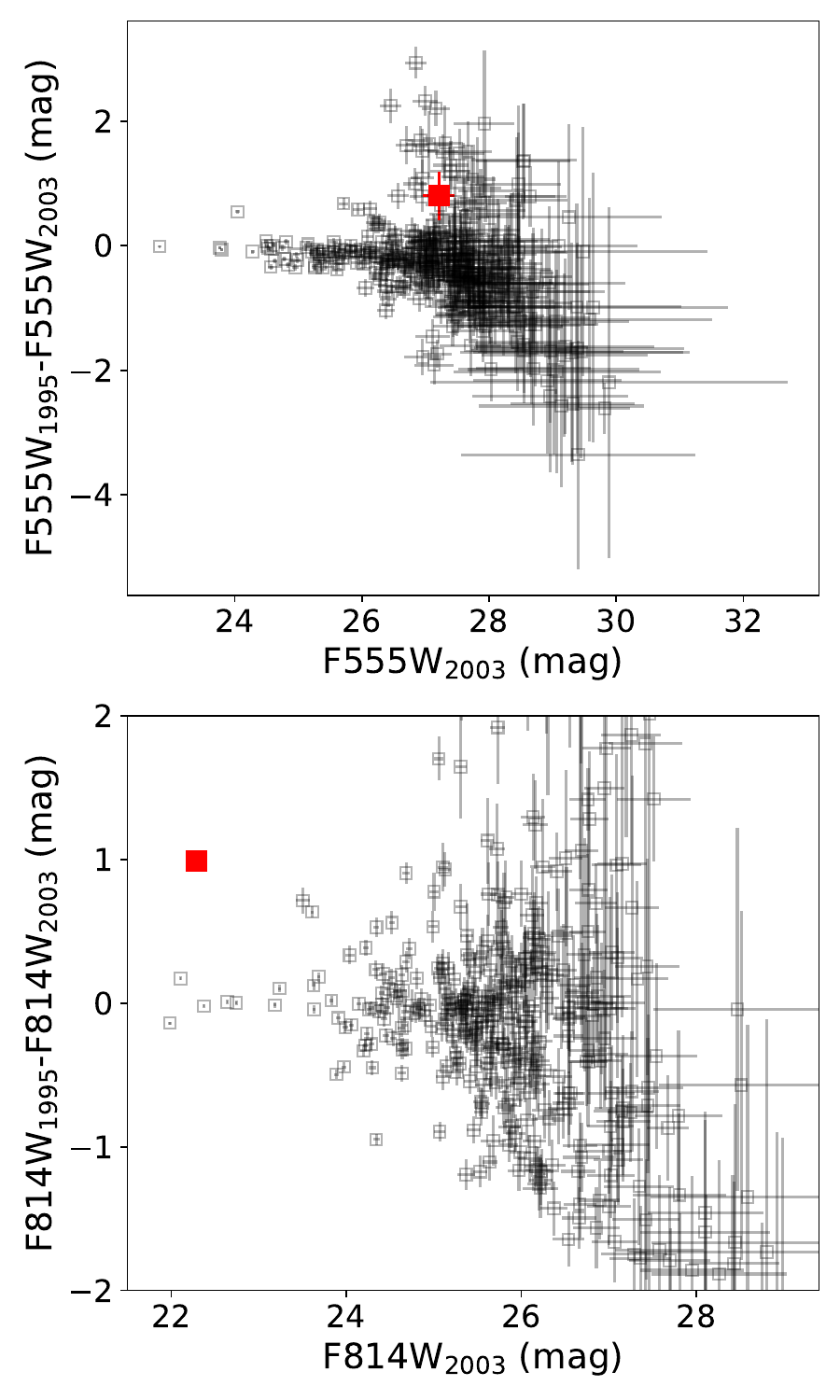}
    \caption{Magnitude changes of the surrounding stars within 5\arcsec\ around the progenitor candidate in F555W (\textit{upper}) and F814W (\textit{lower}) bands from 1995 to 2003. The red point in each panel represents the progenitor candidate of SN~2024ggi.}
    \label{fig:HST_mag_viariation}
\end{figure}

The magnitudes of the progenitor varied by about 1~mag from 1995 to 2003.
To examine whether this variability is intrinsic, we analyzed the properties of the field stars around the SN site.
Fig.~\ref{fig:HST_mag_viariation} shows the magnitude variations of the progenitor candidate and the stars within a radius of 5\arcsec\ and  in F555W and F814W bands. In F555W band, the brightest stars (m$<$26 mag) have little variation, while faint stars (including the progenitor) all show large magnitude changes. Thus, the magnitude change in F555W may not be intrinsic but due to incorrect zeropoints of the two images, or that measurements of these faint objects are less reliable.
In F814W band, the progenitor is one of the two brightest stars, and only one of the other bright stars (m$<$24 mag) also has large magnitude change. Thus, the variability seen for the progenitor candidate in F814W and MIR bands should be true.

\section{Estimation of contribution of nearby stars in infrared bands} \label{appendix:field-stars}
The \textit{Spitzer}/IRAC MIR images and VHS NIR images have resolutions much lower than that of the \textit{HST} images. 
Photometry was done with an aperture radii of 1.2\arcsec\ and 1.0\arcsec\ on \textit{Spitzer}/IRAC and VHS images, respectively.
Besides the progenitor candidate, several other bright stars are found within the photometric aperture of the IR images. Thus, contribution from other stars which were included in the aperture of IR images should be subtracted from the measured IR fluxes for the progenitor.

\begin{figure}
    \centering
    \includegraphics[width=0.9\linewidth]{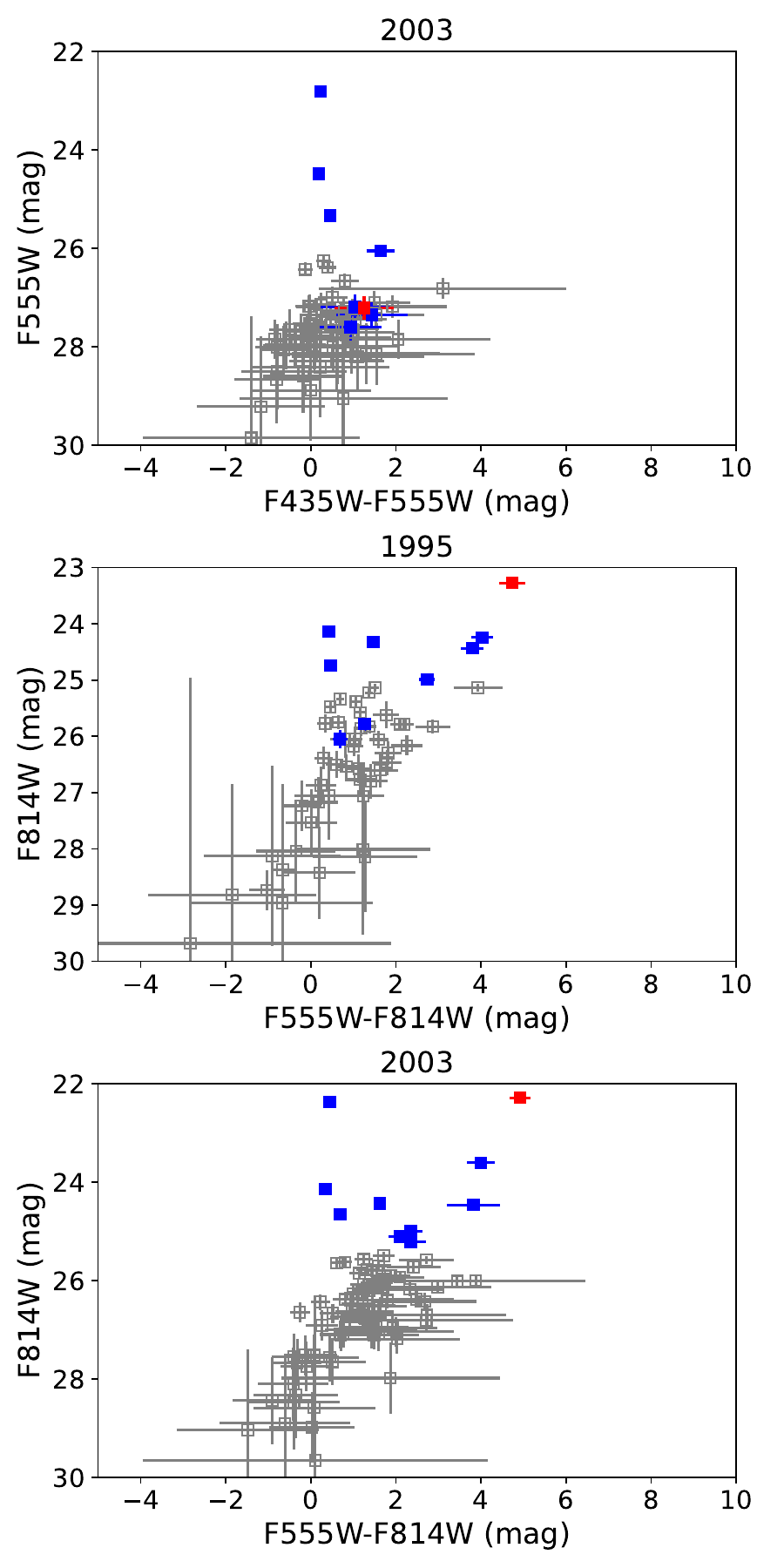}
    \caption{The observed colors and magnitudes of stars within 1.2\arcsec\ around the progenitor candidate of SN~2024ggi. The blue and black points are field stars, and the red points are the progenitor candidate in each panel. \textit{Upper}: F435W$-$F555W \textit{vs.} F555W on the combined 2003 image of HST; \textit{Middle}: F555W$-$F814W \textit{vs.} F814W on the combined 1995 image; \textit{Lower}: F555W$-$F814W \textit{vs.} F814W on the combined 2003 image.}
    \label{fig:HST-star-CM}
\end{figure}

As shown in Fig.~\ref{fig:HST-star-CM}, the progenitor is the brightest in F814W-band (marked by blue squares). The magnitudes of these stars are listed in Tab.~\ref{tab:HST-fieldstars}. Among these bright stars, stars 1, 3 and 6 are quite blue, so we ignored their contribution in IR bands. Stars 2 and 5 are fainter than the progenitor and have slightly bluer colors. Thus, for these two stars, we assume that contributions of the IR fluxes to their total fluxes should not be larger than that of the progenitor. 
In other words, the magnitude difference between the two stars and the progenitor in IR bands should not be less than the that in the F814W band.
Star 2 is 1.3~mag fainter than the progenitor in F814W. That means, the flux of star 2 is less than $\sim$30\% that of the progenitor in IR band. Similarly, star 5 contributes to the emission of the progenitor by $\lesssim$14\% in IR bands. Other stars are too faint so that they contribute little to the total IR fluxes. 
In conclusion, in CH1 and CH2 bands, the progenitor's contribution is 70\%--100\%, and in J and Ks bands, the progenitor's contribution is 88\%--100\%, since star 2 is not included in the VHS aperture.

\begin{figure}[h]
    \centering
    \includegraphics[width=0.5\linewidth]{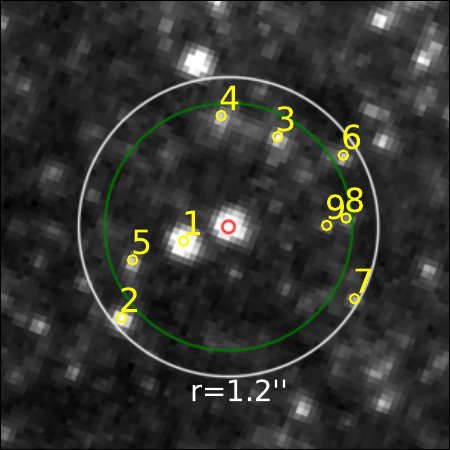}
    \caption{Field around the progenitor candidate (marked by red circle) on the ACS-F814W image in 2003. The yellow circles mark the brightest 9 stars, and the red circle marks the progenitor. The white circle has a radius of 1.2\arcsec\ centered on the progenitor, while the green circle has a radius of 1\arcsec.}
    \label{fig:HST-stars-img}
\end{figure}

\begin{table}
    \centering
    \caption{Magnitudes and 1-$\sigma$ error of the bright stars on the \textit{HST}/ACS pre-explosion images in 2003. Stars are sorted by the magnitude in F814W. Positions of the stars are shown in Fig.~\ref{fig:HST-stars-img}.}\label{tab:HST-fieldstars}
    \begin{tabular}{cccc}
    \toprule
    Number &F435W  &F555W  &F814W\\
    \hline
    progenitor &28.46$\pm$0.66 &27.21$\pm$0.24 &22.29$\pm$0.01\\
    1 &23.04$\pm$0.01 &22.82$\pm$0.01 &22.37$\pm$0.01\\
    2 &28.54$\pm$0.66 &27.61$\pm$0.32 &23.61$\pm$0.02\\
    3 &24.68$\pm$0.03 &24.49$\pm$0.03 &24.14$\pm$0.03\\
    4 &27.69$\pm$0.32 &26.05$\pm$0.09 &24.43$\pm$0.04\\
    5 &- &28.29$\pm$0.63 &24.46$\pm$0.04\\
    6 &25.79$\pm$0.07 &25.34$\pm$0.06 &24.65$\pm$0.04\\
    7 &28.79$\pm$0.81 &27.35$\pm$0.27 &25.00$\pm$0.06\\
    8 &28.24$\pm$0.77 &27.20$\pm$0.25 &25.10$\pm$0.06\\
    9 &- &27.55$\pm$0.36 &25.21$\pm$0.06\\
    \hline
    \end{tabular}
\end{table}

\bibliography{ref}
\bibliographystyle{aasjournal}



\end{document}